\documentclass[12pt]{iopart}
\usepackage{iopams}  
\usepackage{harvard}

\usepackage{epstopdf}

\usepackage{url}
\usepackage[scaled]{helvet}
 
\usepackage[T1]{fontenc}

\usepackage{graphicx}
\newcommand{\egi}{$G\textsubscript{e}$}
\newcommand{\dif}{\mathrm{d}}
\usepackage{enumitem}

\usepackage{color}

\begin{document}

\title{A Gini approach to spatial CO\textsubscript{2} emissions}

\author{Bin Zhou$^{1,2}$, Stephan Thies$^1$, Ramana Gudipudi$^1$, Matthias K.\ B.\ L\"udeke$^1$, J\"urgen P.\ Kropp$^{1,3}$, Diego Rybski$^1$}

\address{$^1$ Potsdam Institute for Climate Impact Research, 14473, Potsdam, Germany}
\address{$^2$ Desert Architecture and Urban Planning Unit, Swiss Institute for Dryland Environmental and Energy Research, Jacob Blaustein Institutes for Desert Research, Ben-Gurion University of the Negev, Israel}
\address{$^3$ Department of Geo- and Environmental Sciences, University of Potsdam, 14476, Potsdam, Germany}
\ead{ca-dr@rybski.de}
\begin{abstract}
Combining global gridded population and fossil fuel based CO\textsubscript{2} emission data at 1\,km scale, we investigate the spatial origin of CO\textsubscript{2} emissions in relation to the population distribution within countries.
We depict the correlations between these two datasets by a quasi-Lorenz curve which enables us to discern the individual contributions of densely and sparsely populated regions to the national CO\textsubscript{2} emissions. 
We observe pronounced country-specific characteristics and quantify them using an indicator resembling the Gini-index. 
Relating these indices with the degree of socio-economic development, we find that in developing countries locations with large population tend to emit relatively more CO\textsubscript{2} and in developed countries the opposite tends to be the case.
Based on the relation to urban scaling we discuss the connection with CO\textsubscript{2} emissions from cities.
Our results show that general statements with regard to the (in)efficiency of large cities should be avoided as it is subject to the socio-economic development of respective countries.
Concerning the political relevance, our results suggest a differentiated spatial prioritization in deploying climate change mitigation measures in cities for developed and developing countries.
\end{abstract}

%\submitto{none}

\maketitle

\section{Introduction}

Urbanization is an ongoing process in many parts on the globe. 
It is projected that due to rural-urban migration much of the future urbanization is going to take place in developing and transition countries leading to ever more mega-cities \cite{UN2015,WiggintonUWM2016}.
In parallel, humanity is facing another challenge, namely climate change. 
To date, cities, despite occupying less than 1\,\% of the global land area, account for more than 70\,\% of the anthropogenic green house gas (GHG) emissions \cite{UNHABITAT2011}. Therefore, cities are often identified as the key for global mitigation actions.  
While a large contribution of the global CO\textsubscript{2} emissions is commonly attributed to cities \cite{Seto2012}, the CO\textsubscript{2} reduction role of further urbanization is also discussed with the argument of efficiency gains associated with the high densities in cities \cite{DodmanD2009}. 
Moreover, cities are known to perform more efficiently in addressing the basic needs of human beings \cite{DodmanD2009}. 
Hence, a diversified view on cities is needed and in view of climate change mitigation, a better understanding of the interplay between urbanization, origin of CO\textsubscript{2} emissions, and socio-economic development is of great interest.

Globally, cities are characterized by higher population densities compared to rural areas.
Recent literature has identified the crucial role played by population density in either increasing or decreasing the emission efficiency in cities \cite{NewmanK1989,GlaeserK2010,BrownSS2009,JonesK2014,KennedySGHHHPPRM2009}. 
The impact of population density on reducing/increasing CO\textsubscript{2} emissions in these studies is either calculated based on specific assumptions made to calculate the city specific CO\textsubscript{2} emissions or through the construction of city clusters using a clustering algorithm, see \cite{OliveiraAM2014,GudipudiFRW2016}. 
However, most of these studies are limited to a specific country or a region. Therefore, there is a gap in the existing literature about the sub-national origin of CO\textsubscript{2} emissions at a global scale. 
Bridging this gap would provide better insights as to whether population density is a crucial factor in improving/decreasing emission efficiency and would identify other factors that influence CO\textsubscript{2} emissions at a country scale.  

Here, we investigate how the spatial origin of CO\textsubscript{2} emissions relates to the spatial distribution of population.
In order to avoid discussions about the proper city definition, the correlations are analyzed on the level of grid cells -- keeping in mind that locations of high population are likely to belong to cities.
Thus, we analyze population and CO\textsubscript{2} emissions by employing a quasi-Lorenz curve that relates the cumulative population and cumulative emissions for entire countries on a grid-cell level (the Lorenz curve was originally used to describe unequal income distribution).
The shape of these curves explains whether the emissions are concentrated in locations of high or low population. 
Inspired by the apparent similarity, we extend the well-known Gini-index.
Based on the data employed, we find that, within many countries, locations with high or low population exhibit different relative emissions. 
We thus compare the extended ``Gini-index'' with the economic strength of the considered countries (as captured by the GDP per capita) which can be to some extent interpreted as a measure for the stage of development. 
We further hypothesize that the development stage of respective countries plays an important role in explaining this relationship. 

Earlier studies attempted to address the emission efficiency of densely populated regions by means of urban scaling, where an urban indicator is plotted against the city size in terms of population \cite{BettencourtLHKW2007}. 
The exponent, estimated as the slope of a linear regression in the log-log representation, 
quantifies efficiency gains of large or small cities.
However, in case of urban CO\textsubscript{2} emissions, published results from urban scaling leave an inconclusive picture [for an overview we refer to \cite{RybskiRWFSK2017,GudipudiRLZLK2017}].
In the present work we address this issue by combining high resolution, global population and CO\textsubscript{2} emission datasets in order to quantify whether locations with high or low population emit more or less CO\textsubscript{2}.
We further discuss an analytical link between our approach and urban scaling.

This paper is organized as follows. Firstly, we describe the main datasets this work is based on, namely the population and the CO\textsubscript{2} emissions data in Sec.~\ref{sec:data}. 
Then, the analyses and results are presented in Sec.~\ref{sec:analyses}, which is organized into 5 sub-sections, i.e.\ describing the employed index, the comparison among and within countries, and a robustness analysis.
It also contains the connection to urban scaling.
Finally, we summarize and discuss our work in Sec.~\ref{sec:discussion}.
Further details are included in the Appendix.

\section{Data}
\label{sec:data}
\subsection{Population data}
\begin{figure}
\centering
\includegraphics[width=\linewidth]{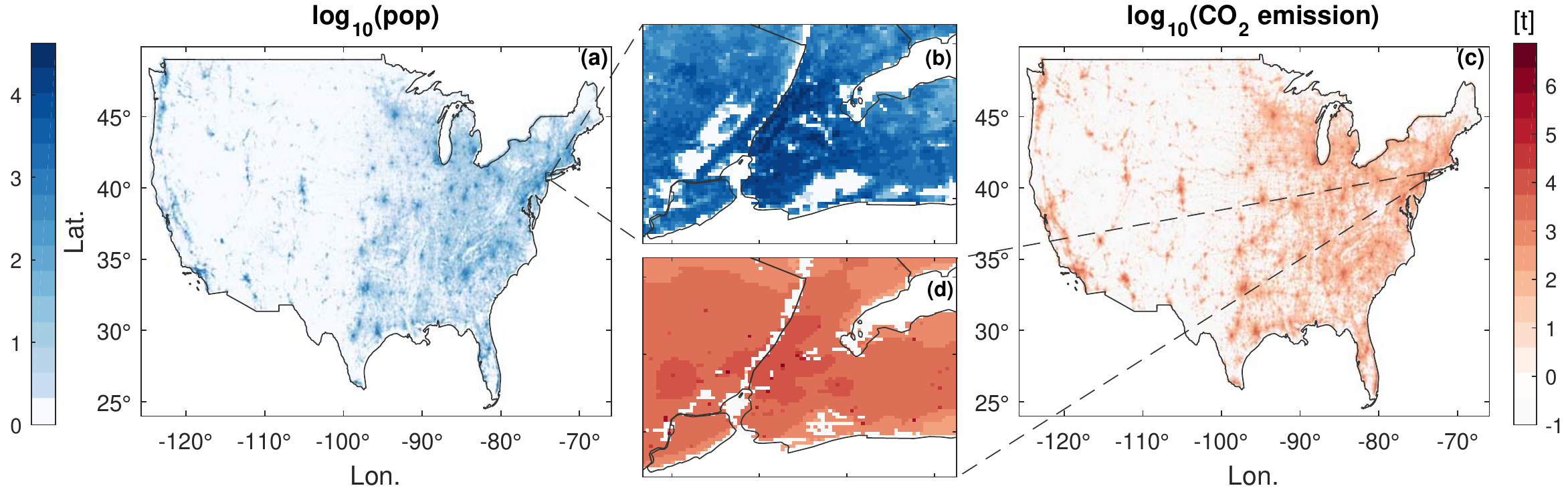}
\caption{Population and CO\textsubscript{2} emissions for the contiguous USA. (a) Gridded Population of the World, version 4, GPWv4 \cite{WhitfieldMAP2015} at $1\times 1$\,km$^2$ spatial resolution in 2010 and (c) total anthropogenic CO\textsubscript{2} emission from ODIAC data \cite{OdaM2011} for the same region, year, and resolution. (b) and (d) depict magnified views of population and CO\textsubscript{2} emission for the New York metropolitan region, respectively. Visually, large agglomerations of population coincide with large amounts of emissions. To which extent they relate proportionally is the subject of this paper.}
\label{fig:grump_odiac}
\end{figure}

We used the Gridded Population of the World, version 4 (GPWv4) population count data for the year 2010 \cite{CIES2016}. GPWv4 data allocate the population counts of census units collected globally from various institutions into standard $1\times 1$\,km$^2$ grid cells by means of an areal-weighting interpolation \cite{WhitfieldMAP2015}.
Figure~\ref{fig:grump_odiac}(a) illustrates the GPWv4 data in the year 2010 for the contiguous US. The distribution of population in the US exhibits an inhomogeneity.
The metropolitan urban agglomerations accommodate a large share of population in the US, whereas the states in the Mountain West are generally sparsely populated.

\subsection{CO\textsubscript{2} emissions data}
\label{sec:co2data}
Fossil fuel based CO\textsubscript{2} emission estimates are obtained from the Open source Data Inventory of Anthropogenic CO\textsubscript{2} (ODIAC) emissions available globally at $1\times 1$\,km$^2$ grid \cite{OdaM2011}. In the ODIAC dataset, point sources, i.e.\ power plant emissions obtained from the database CARMA (Carbon Monitoring and Action) are directly assigned to the grids, while non-point sources (e.g.\ emissions from transport, industrial, residential, and commercial sectors) 
are disaggregated based on the national and regional emission estimates, using remotely sensed nightlight data.
In order to ensure comparability of fossil fuel CO\textsubscript{2} emissions across the globe and being restricted to the data availability, the ODIAC dataset does not include the emissions from cement production. Non-land emissions, such as those from international bunkers (international aviation and maritime shipping), are assigned to the non-point emissions.

Compared with conventional population-based approaches, the nightlight data can trace the human activities more appropriately \cite{ElvidgeBDB1999,ElvidgeIBH2001}. 
Worthy of special mention is that the gridded emission data are not disaggregated using population density as a proxy. Therefore, the two datasets depict distinguishing zonal patterns, as shown in the example for the New York metropolitan region in Fig.~\ref{fig:grump_odiac}(b) and~(d).
Without relying on the time-consuming update of census data, emissions allocated using nightlights can be updated more frequently and may be of particular importance for developing countries where conducting census is still a challenge.
Figure~\ref{fig:grump_odiac}(c) shows the gridded total anthropogenic CO\textsubscript{2} emissions (in tons) for the year 2010 for the contiguous US, analogous to the population data shown in Fig.~\ref{fig:grump_odiac}(a). 
As can be observed, the emissions also exhibit pronounced inhomogeneities. 

We compare our results obtained from the ODIAC data with other CO\textsubscript{2} emission datasets, namely the Fossil Fuel Data Assimilation System (FFDAS) and the Emission Database for Global Atmospheric Research (EDGAR).
For the sub-national analysis we also analyze the Vulcan data, which has been analyzed before \cite{GudipudiFRW2016}.
However, we focus on ODIAC, since it has highest resolution, and we discuss the other datasets in comparison.

The fundamentals of creating the four gridded CO\textsubscript{2} emission inventories have been compared and discussed in detail in \cite{HutchinsCMM2016}. In general, they differentiate themselves in terms of 1) the energy statistics used which determines the sectors included in calculating the total national CO\textsubscript{2} emissions, and 2) the approach to disaggregating and allocating the CO\textsubscript{2} emissions to a regular grid.

Dissimilarities among the inventories may be dominated by the disaggreation method.
FFDAS applies the Kaya identity to balance CO\textsubscript{2} emssions across regions, relying on population and nightlight data \cite{RaupachRP2010} [see also \cite{GudipudiRLZLK2017} for further information on the Kaya identity]. 
Viewed as the most accurate emissions inventory, a bottom-up method has been used for the Vulcan data allocate large point sources, road-specific emissions, and non-point emissions to census tracts, and further resampled to a 10-km grid \cite{GurneyMZF2009}. However, since the subnational emissions data are not always available, the Vulcan data is restricted to the USA at the moment.

\section{Analyses and results}
\label{sec:analyses}

\subsection{Inhomogeneity of emissions and the index \egi} 
In order to characterize the relation between country-wise population and CO\textsubscript{2} emissions, we plotted the cumulative quantities against each other.
Therefore, we sorted the grid cells of a country by population in ascending order and calculated the cumulative share of population and CO\textsubscript{2} emissions arising therefrom.
Then we plotted the cumulative emissions as a function of the corresponding cumulative population.

\begin{figure}
\includegraphics[width=\linewidth]{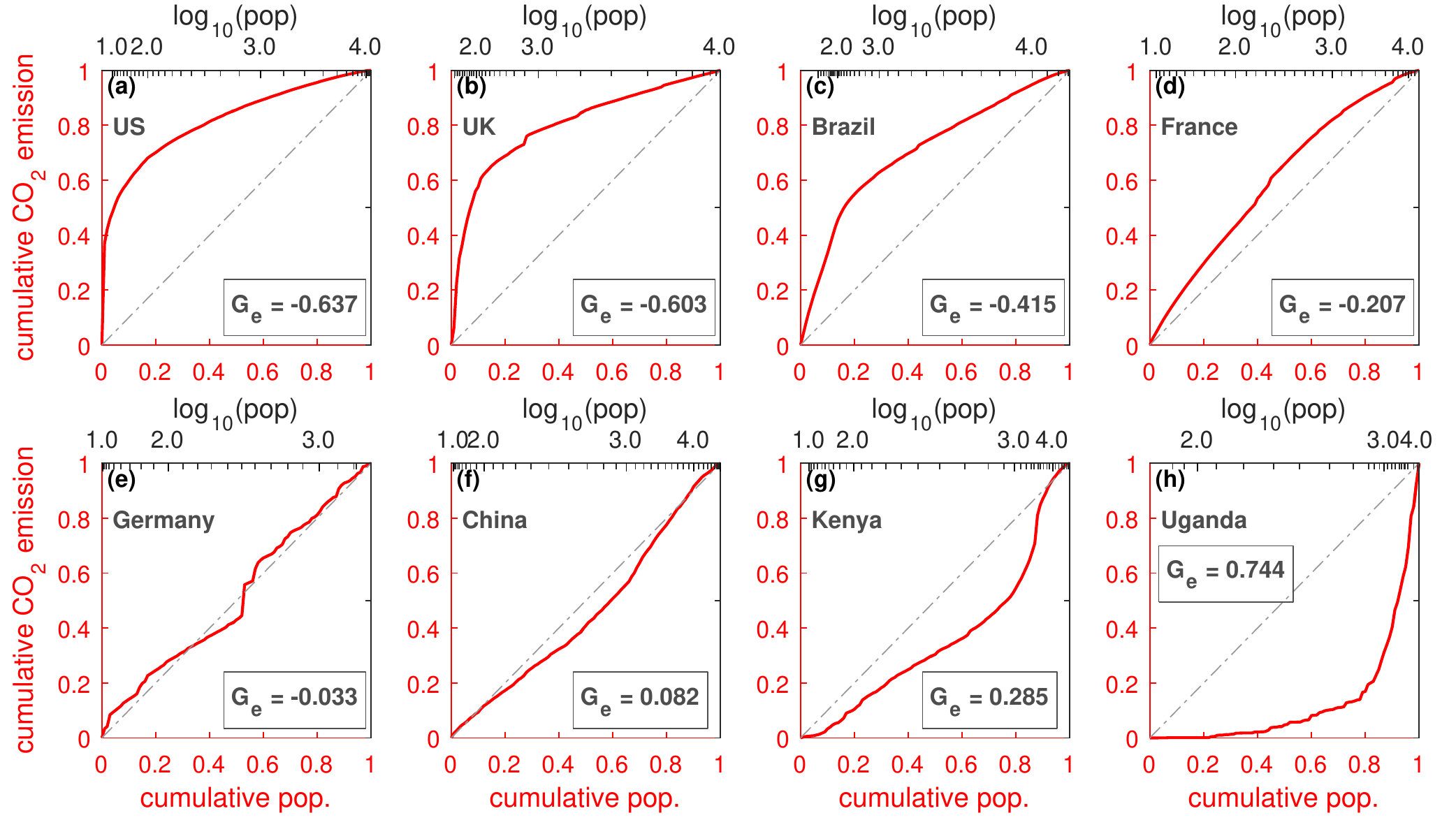}
\caption{Quasi-Lorenz curves and corresponding inhomogeneity index \egi{} for selected countries. The country-specific curves are drawn by plotting the accumulated population (in ascending order) on the horizontal axis against the accumulated share of CO\textsubscript{2} emissions of the corresponding grid cells. The panels show the curves for (a)~USA, (b)~UK, (c)~Brazil, (d)~France, (e)~Germany, (f)~China, (g)~Kenya, and (h)~Uganda. If the curves follow the diagonal, then low and high densities have the same emissions per capita. If the curves are bent to the lower right corner, then cells of small density exhibit relatively low emissions and high population cells exhibit relatively high emissions. Curves in the upper left corner indicate the opposite behavior. The inhomogeneity index \egi{} is positive or negative, respectively. It can be seen, that various countries exhibit non-proportional relations between population and emissions. The inhomogeneity index \egi{} seems to be related to the development of the country.}
\label{fig:gcurves}
\end{figure}

Figure~\ref{fig:gcurves} shows the resulting curves for a few countries.
In the case of constant emissions per capita, the two cumulative quantities would be proportional to each other and follow the diagonal in Fig.~\ref{fig:gcurves}. 
For Germany, Fig.~\ref{fig:gcurves}(e), this is approximately the case, but for other countries the graphs are more or less curved, bent either to the upper left or to the lower right corner.

Curves like the ones displayed in Fig.~\ref{fig:gcurves} resemble the so-called \emph{concentration curves} used to describe socio-economic inequalities. The most popular concentration curve is the Lorenz curve usually employed to visualize income inequalities. Other applications of concentration curves include for example the analysis of socio-economic inequalities in the health sector [e.g.\ \cite{KakwaniWD1997}]. Since the curves we compute here, do not agree exactly with the classical definition of a concentration curve we will refer to them as \emph{quasi-Lorenz curves}.

The slopes of the curves depend on per capita emissions.
If the curves are bent to the upper left corner, as e.g.\ in the case of the USA, Fig.~\ref{fig:gcurves}(a), or the UK, Fig.~\ref{fig:gcurves}(b), then the grid cells with small population already include a large amount of emissions (large slope). 
On the contrary, if the curves are bent to the lower right corner, as e.g.\ in the case of Uganda, Fig.~\ref{fig:gcurves}(h), or Kenya, Fig.~\ref{fig:gcurves}(g), then many grid cells with small population are necessary to include a fair amount of emissions (small slope).
Accordingly, curves bent to upper left indicate high per capita emissions in sparsely  populated cells and comparably lower per capita emissions in densely populated cells, and vice versa.

In Germany, per capita CO\textsubscript{2} emissions of large cities are smaller than those of small ones, but the difference seems to be minor \cite{SchubertWG2013}.
In contrast, per capita CO\textsubscript{2} in the UK emissions remarkably diverge between large and small cities, ranging from 25.6 tonnes per capita in Middlesbrough to 5.4 tonnes per capita in London in 2012, reflecting the impact of industrial base
\cite{CC2015,DECC2016}.

\begin{figure}
\centering
\includegraphics[width=0.45\linewidth]{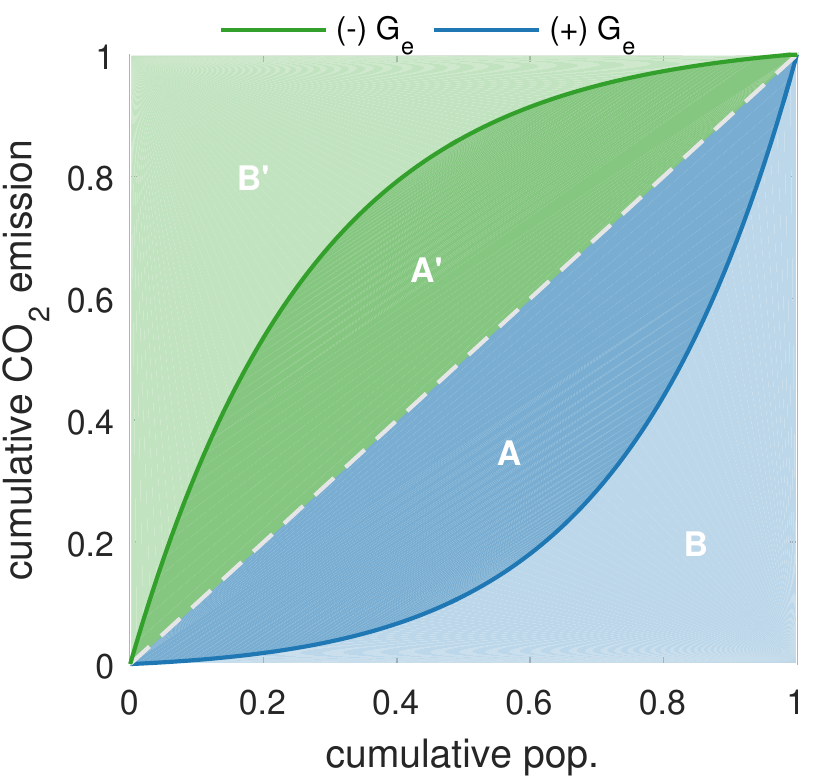}
\caption{Illustration of the inhomogeneity index \egi{}. Quasi-Lorenz curves (solid lines) and the calculation of \egi{}, which is inspired by the Gini coefficient: \egi$_+$  =  $A /(A+B)$ and \egi$_-$  =  $-A' /(A'+B')$.
}
\label{fig:def_ge}
\end{figure}

Interestingly, in Fig.~\ref{fig:gcurves} developed countries seem to belong to the group where the curves extend to the upper left corner and less developed countries seem to belong to the group where the curves extend to the opposite corner. 
In order to verify if this is a coincidence or if there is a systematic relation, we first need to break the shape of the curves down to a single number.
Therefore, we generalize the Gini coefficient, which originally has been introduced to quantify income inequality \cite{GiniC1921}. 
As illustrated in Fig.~\ref{fig:def_ge}, we distinguish between curves above or below the dashed line with a slope of 45$^\circ$ -- the line of equality. For the blue quasi-Lorenz curve, the index is therefore defined as the ratio of the area between the curve and the line of equality (marked as $A$) to the total area below the line of equality ($A+B$). Analogously, the index of the green curve is $-A'/(A'+B')$. We arbitrarily assign the index for the curves above the line of equality negative, and below positive.

\subsection{\egi{} versus GDP per capita at trans-national level} 
\begin{figure}
\centering
\includegraphics[width=0.5\linewidth]{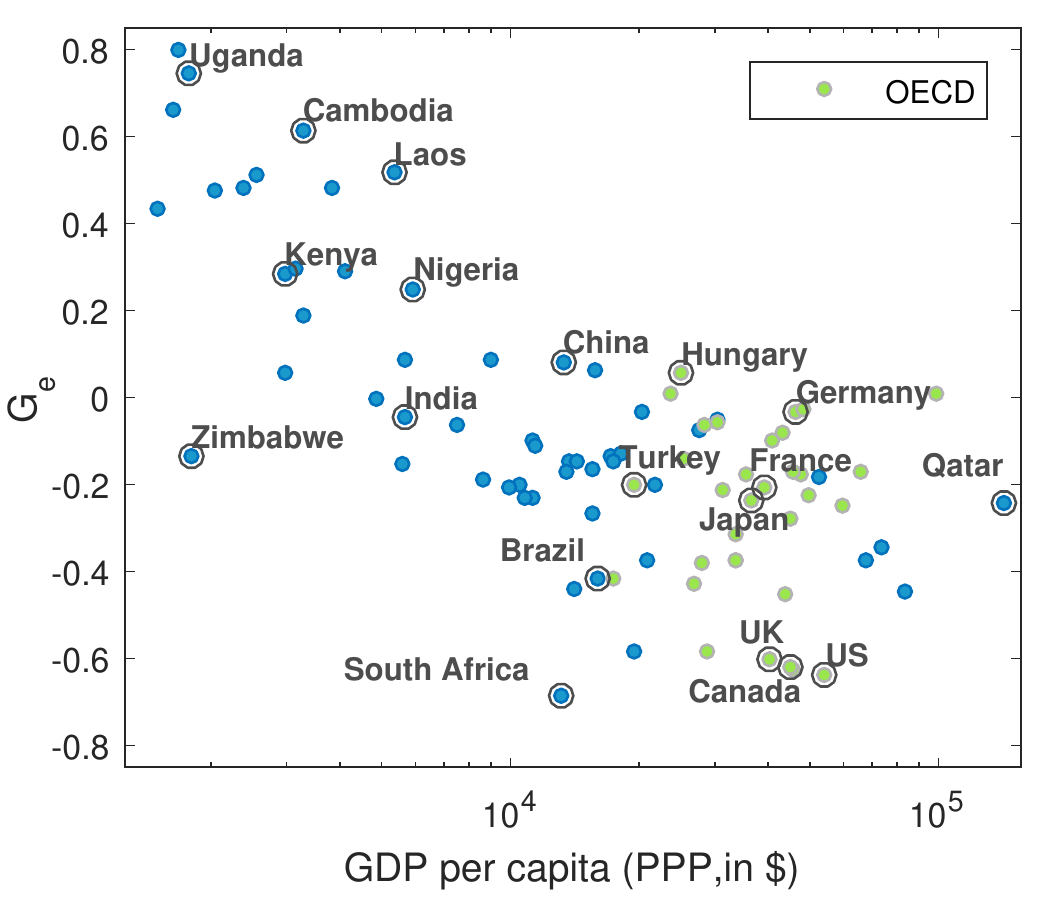}
\caption{Development dependence of CO\textsubscript{2}-population-inhomgeneity.
The inhomogeneity index \egi{} is plotted vs.\ Gross Domestic Product (GDP) per capita (PPP) for 94 countries on a semi-logarithmic scale.
For better readability only the symbols of a sub-set of countries are labeled. 
As can be seen, the \egi{} correlate with the economic development. 
The Pearson correlation coefficient between \egi\ and GDP on a logarithmic scale is $\rho=-0.71$ ($p\le 0.01$).
In more developed countries high population densities have lower emissions as low densities.
The GDP data were obtained from the World Bank (\url{http://data.worldbank.org}), measured in USD of the year 2010.}
\label{fig:ge_vs_gdp}
\end{figure}

In Fig.~\ref{fig:ge_vs_gdp} the values of the inhomogeneity index \egi{} for a large number of countries are plotted against the logarithm of GDP Purchasing Power Parity (PPP) per capita obtained from the World Bank, an important indicator for economic development.
As observed, the two quantities correlate (with a Pearson correlation coefficient $\rho = -0.71$, $p\le 0.01$).
In general, for developed countries \egi{} tends to have smaller values, and for developing ones it tends to have larger values.
Thus, we generalize that in economically developing countries, high population densities are more emission intense and the opposite is the case in economically developed countries.

\begin{figure}
\centering
\includegraphics[width=0.75\linewidth]{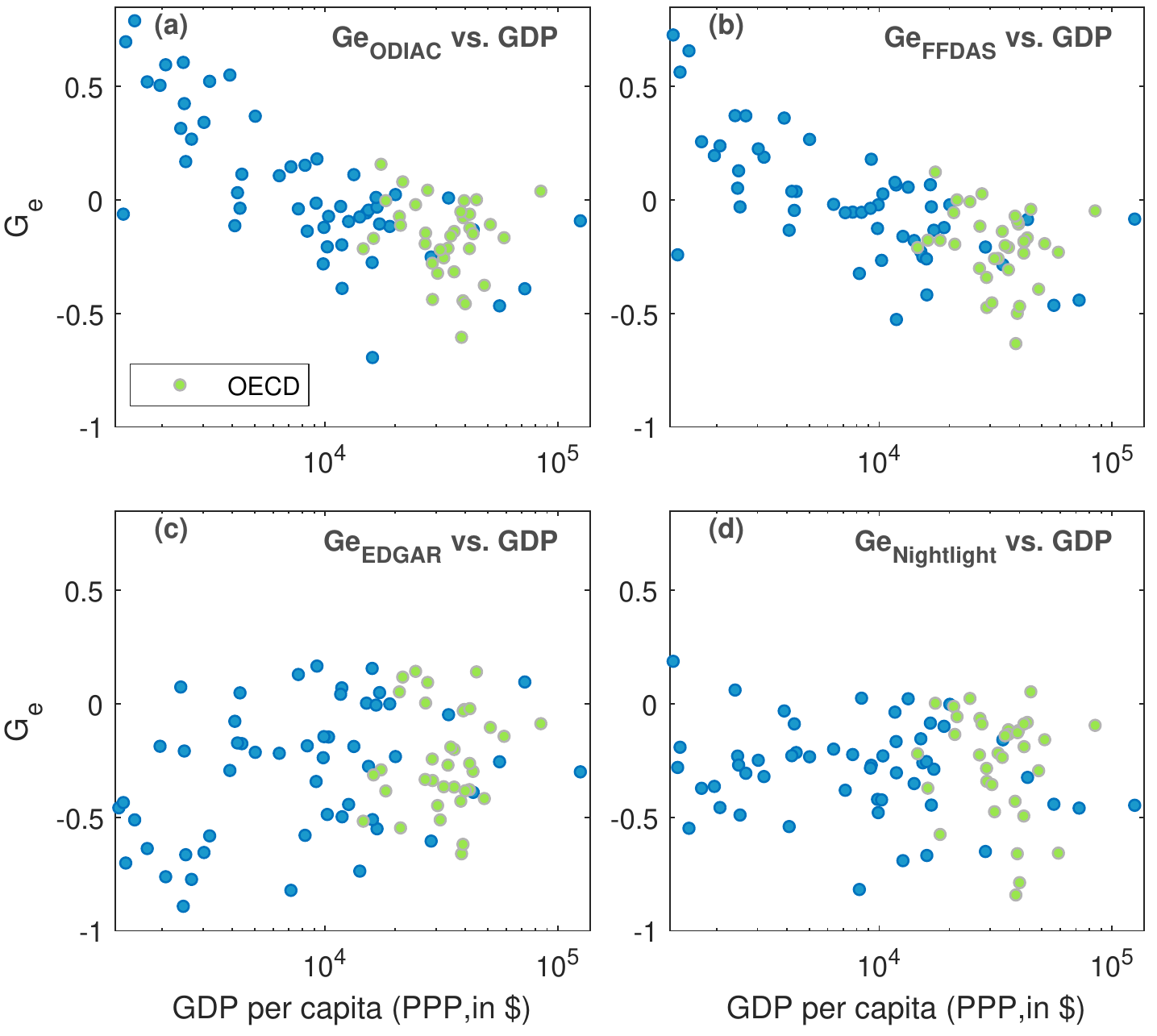}
\caption{Comparison of CO\textsubscript{2}-population-inhomgeneity for different CO\textsubscript{2} datasets and nightlights.
(a) ODIAC, (b) FFDAS, (c) EDGAR, (d) nightlights \cite{ElvidgeBZH2017}.
Each panel is analogous to Fig.~\ref{fig:ge_vs_gdp} but for consistency of spatial resolution the underlying ODIAC data in (a) has been aggregated to 10\,km resolution.
}
\label{fig:comp4}
\end{figure}

We repeat the analysis for the FFDAS and the EDGAR data (Sec.~\ref{sec:co2data}).
For reasons of consistency, we also analyze the ODIAC data aggregated to 10\,$\times$10\,km$^2$ resolution.
For the three datasets, in Fig.~\ref{fig:comp4} the resulting \egi{}-values are plotted against the GPD per capita, analogous to Fig.~\ref{fig:ge_vs_gdp}.
As can be seen, for FFDAS [Fig.~\ref{fig:comp4}(b)] a very similar development dependence as for ODIAC [Fig.~\ref{fig:comp4}(a)] is found. 
In contrast, for the EDGAR data [Fig.~\ref{fig:comp4}(c)] the development dependence vanishes and is even slightly inverted (\textbf{$\rho = 0.29$, $p \le 0.01$}).
Differences between the \egi{}-values of the EDGAR and ODIAC or FFDAS data are most pronounced for developing countries. 
We are unclear about the reasons for the different outcomes.
Reasons could be the poor quality of population census, high demographic dynamics, and insufficient geo-spatial data in developing countries. 
EDGAR relies on road networks, population density and agriculture land use data to downscale the national emissions, which renders it more sensible to errors embedded in the proxy datasets. 
Moreover, the EDGAR database includes emissions from cement production. 
In comparison, FFDAS uses, besides population density, nightlight data to disaggregate emissions. 

Moreover, since ODIAC and FFDAS are at least partly based on nightlight data for the subnational disaggregation \cite{OdaM2011,RaynerRPP2010,RaupachRP2010}, one may argue that the development dependence in Fig.~\ref{fig:comp4}(a) and~(b) are simply due to such an effect in the nightlight data.
Thus, we also analyzed the nightlight data from the Visible Infrared Imaging Radiometer Suite (VIIRS) Day/Night Band (DNB) data \cite{ElvidgeBZH2017} in an analogous way as the emissions data and the results are displayed in Fig.~\ref{fig:comp4}(d).
As can be seen, for nightlight data, we do not see any correlations between \egi{}-values and the GDP per capita.
Accordingly, we conclude that the development dependence found in the ODIAC and FFDAS data is not stemming from the nightlight data.
Overall, \egi{}-values tend to be negative for nightlights which indicates that locations of low population have a relatively strong contribution.

\subsection{\egi{} versus GDP per capita at sub-national level}  
\begin{figure}
\includegraphics[width=\linewidth]{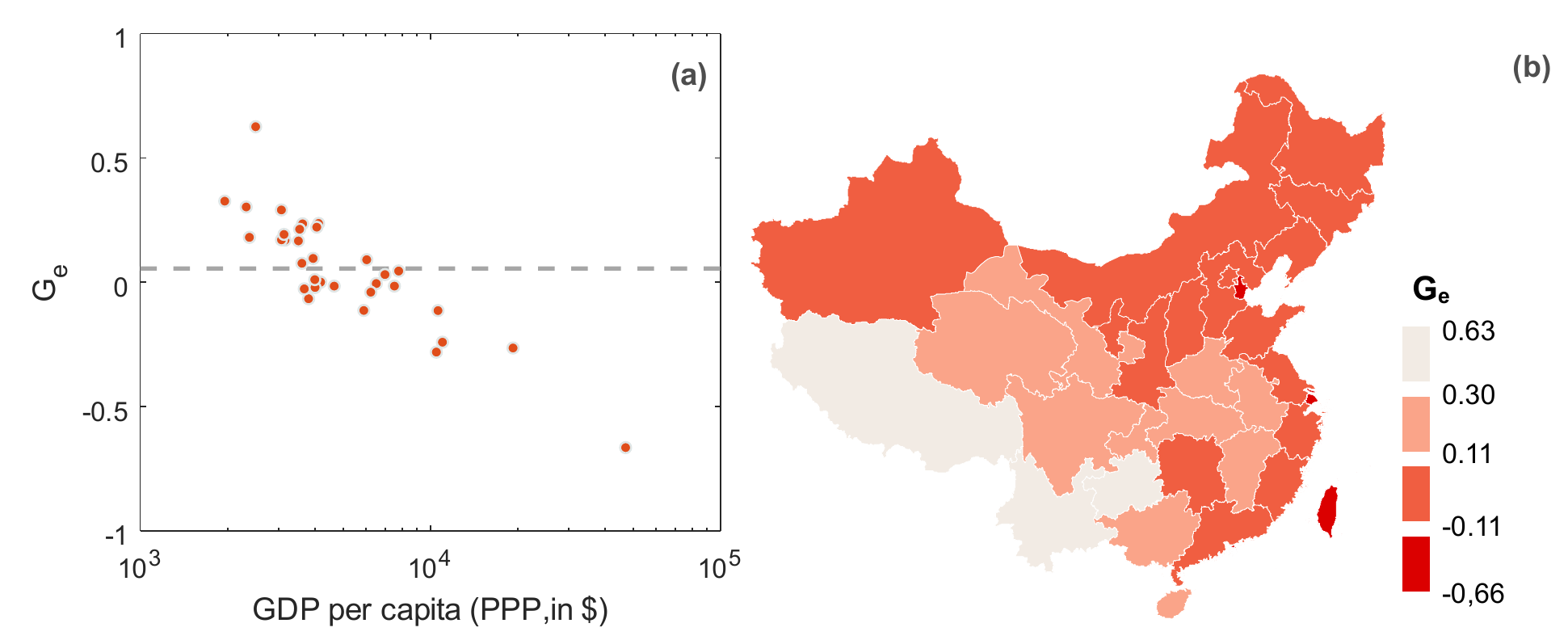}
\caption{Sub-national inhomogeneity index \egi{}. 
We calculated the \egi{} on the province level for China. 
In (a) the \egi{}-values are plotted against the corresponding province GDP per capita values on a logarithmic scale, analogous to Fig.~\ref{fig:ge_vs_gdp}.
The dashed line indicates the country-level mean \egi.
Panel (b) shows a map of China where the provinces are color-coded according to the inhomogeneity index \egi{}. 
It can be seen that the development dependence as found in Fig.~\ref{fig:ge_vs_gdp} does also hold on the sub-national scale -- at least in China.
Provinces with lowest and highest \egi{}-values are Hong Kong and Tibet, respectively.
Note, however, that for the USA we do not find sub-national correlations (\ref{sec:subnatusa}).}
\label{fig:china}
\end{figure}

Next we want to analyze if the correlations between \egi{} (for ODIAC) and GDP per capita among countries also appear within a country.
Therefore, we disaggerate the data of China into provinces. 
Analogously as for the countries, we calculate cumulative emissions vs.\ cumulative population and determine the inhomogeneity index at the province level.
In Fig.~\ref{fig:china}(a) the \egi{}-values are plotted vs.\ the corresponding GDP per capita values, as in Fig.~\ref{fig:ge_vs_gdp} but now for provinces.
Similar to the country analysis and even more pronounced, we find correlations ($\rho=-0.87$, p-value: $<0.01$, statistically significant).

However, performing the corresponding sub-national analysis for the USA on the state level and we could not find significant correlations (\ref{sec:subnatusa}). 
Despite this lack of correlations, we find a spatial pattern in the USA. 
States at the west coast and in the Northeast tend to have larger \egi{}-values.
This is also the case for other states at the east coast and in the Midwest.
States in the south as well as Montana, North Dakota, South Dakota tend to have more extreme \egi{}-values. 
Repeating the analysis for the Vulcan data (Sec.~\ref{sec:data}), which might be considered the most detailed data, still no correlations between \egi{} and GDP per capita within the USA are found (\ref{sec:subnatusa}).
However, the analysis does show weak correlations between the \egi{}-values of Vulcan and ODIAC data.

\subsection{Robustness of \egi{}}
\begin{figure}[htp!]
\centering
\includegraphics[width=0.9\linewidth]{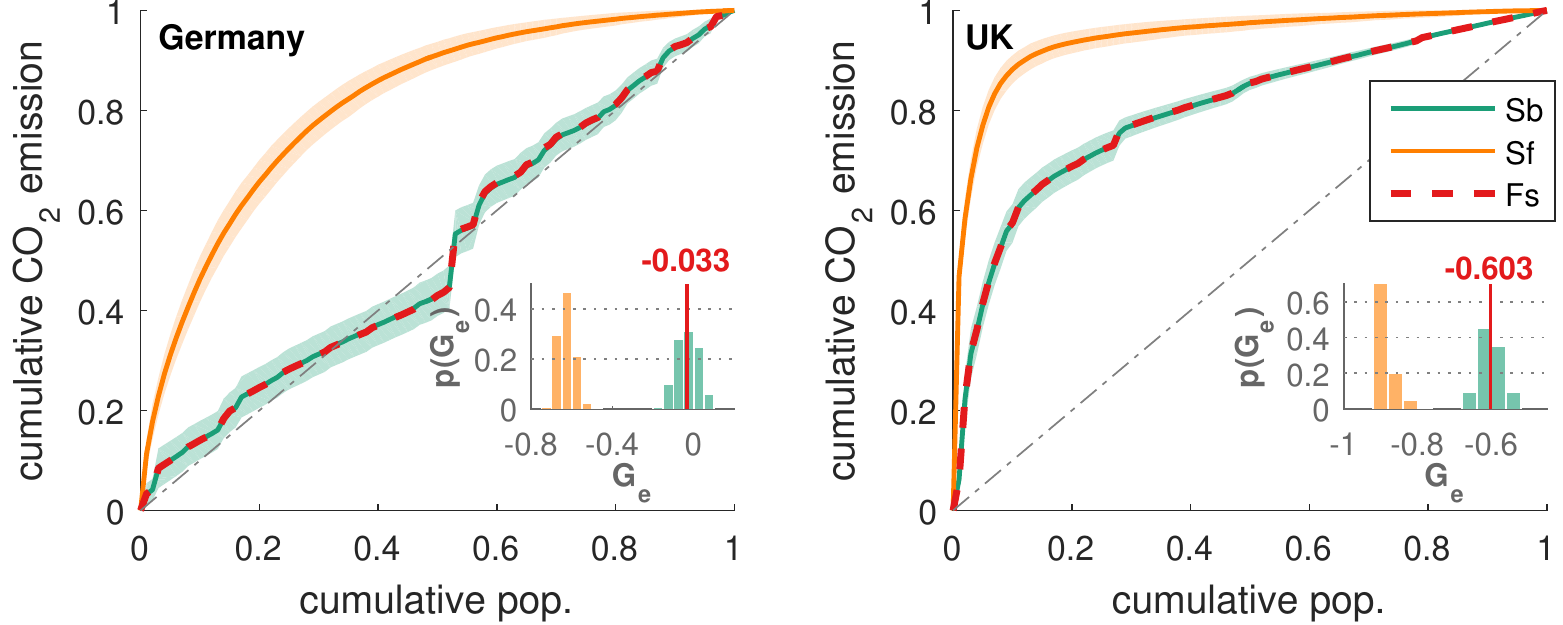}
\caption{Robustness of \egi{}. In order to illustrate the robustness of the curves in Fig.~\ref{fig:gcurves} we compare them with curves when the data is sub-sampled or shuffled. 
The panels for (a) Germany and (b) UK include the curves for the full samples (Fs), median and envelop for the sub-sampled data (Sb, green), and the median and envelop for the shuffled data (Sf, orange). The insets show histograms of the corresponding inhomogeneity indices.
It can be seen that sub-samples of the data lead to similar results as for the full sample so that the results are not due to individual pixels. 
The \egi\ values from the shuffling approach $-1$, as the correlations between population and CO\textsubscript{2} are destroyed.}
\label{fig:robust}
\end{figure}

Last we want to check the robustness of the \egi{} coefficient.
Therefore, we explore different forms of sampling and randomization.
In order to check the influence of outliers, we create random sub-samples of the ODIAC data.
We constructed a set with 50\,\% of the original size by randomly selecting pairs of population and emissions values from the original set without replacement for 1000 iterations. 
We calculated the cumulative quantities as before and determined the inhomogeneity index. 
Repeating the procedure we can assess the statistical spread.
As observed in Fig.~\ref{fig:robust}, the resampling has minor influence on the shape of the curve and the resulting \egi{}-values.
For Germany, Fig.~\ref{fig:robust}(a), 95\,\% of the realizations lead to \egi{}-values in the range of -0.131 to 0.063, with a median and mean of -0.032, which is very close to the measured value -0.033. The sub-sampled robustness check for the UK led to analogous findings [Fig.~\ref{fig:robust}(b)].

Another way to randomize is to shuffle.
Since in the analysis we have already sorted the data, we now shuffle only the emissions data and destroy the correlations between emissions and population.
Then we perform the whole analysis and obtain cumulative emissions and population curves as well as \egi{}-values.
Repeating the procedure we can assess the statistical spreading.
The results are also displayed in Fig.~\ref{fig:robust}, and we find that the curves for the shuffled data are very different from the original curves which shows that the actual shapes in Fig.~\ref{fig:gcurves} are due to the correlations between emissions and population.
The shape of the curves for the shuffled data differs between Germany and the UK, Fig.~\ref{fig:robust}(a) and~(b).
Since shuffling destroys any correlations, the actual form of the curves can be attributed to a combination of the probability distributions of the population and emissions which differ among the countries.

\subsection{Relation to urban scaling}
\label{ssec:scaling}
The analysis of CO\textsubscript{2} efficiency that is carried out here using quasi-Lorenz curves can be related to the urban scaling approach as advocated in \cite{BettencourtLHKW2007}. The urban scaling approach aims to establish a parametric relationship between the urban population  $P_\textrm{u}$ of a city and the respective emissions $E_\textrm{u}$. In our analysis we do not analyze urban population and urban emissions explicitly but examine gridded population $P_\textrm{g}$ and emission data $E_\textrm{g}$ within countries. Since urban areas are usually characterized by high population densities (depending on the pixel size), one could transfer the idea of urban scaling to our setting and assume the scaling relationship $E_\textrm{g} \sim P_\textrm{g}^\beta$. The case of $\beta<1$ indicates CO\textsubscript{2} efficiency gains with increasing population (density) while $\beta>1$ is associated with efficiency losses. 
Here it is of interest how the non-parametric quasi-Gini coefficient \egi{} is related to the parametric scaling exponent $\beta$.

Generally there is no simple association between $\beta$ and \egi. Empirically, the $\beta$ coefficient is usually estimated as the slope of a linear regression of the logarithmic quantities. 
Hence, it depends on the correlations among the logarithmic quantities $\mathrm{cor}\{\log P_\textrm{g}, \log E_\textrm{g}\}$ and on the variance of $\log P_\textrm{g}$ and $\log E_\textrm{g}$ only. 
By contrast, \egi{} as a non-parametric estimator depends on the exact form of the joint distribution of $P_\textrm{g}$ and $E_\textrm{g}$. However, it is possible to determine a specific expression for the relationship between \egi{} and $\beta$ under certain conditions.
The coefficients are related via \cite{Pfaehler1985}
\begin{equation}
G_e=\frac{\beta-1}{2\lambda-\beta-1}
\, ,
\label{equ:betagini}
\end{equation}
if $P_\textrm{g}$ is Pareto distributed with shape parameter $\lambda > 1$ and a scaling relation of the form $E_\textrm{g} \sim P_\textrm{g}^\beta$ with $\beta<\lambda$ holds \emph{exactly}.
For a detailed derivation see \ref{sec:betagini}. 
The formula shows that a scaling coefficient $\beta>1$ is associated with $G_e>0$. Equivalently, $\beta<1$ implies $G_e<0$. If $E_\textrm{g} \sim P_\textrm{g}^\beta$ holds only approximatively, Eq.\ (\ref{equ:betagini}) should still give a reasonable approximation.

Under this scenario, our finding of development dependent $G_e$-values implies a corresponding development dependence of the scaling exponent $\beta$.
Accordingly, in developing countries large cities are typically less emission efficient and vice versa in developed countries.

\section{Summary and Discussion}
\label{sec:discussion}

In summary, we have analyzed the correlations between the spatial distribution of population with CO\textsubscript{2} emissions using high resolution datasets.
In order to understand these correlations we employed the quasi-Lorenz curve.
The shape of the curve indicates to which extent locations of high or low population emit relatively more or less CO\textsubscript{2}.
We characterized the inhomogeneity by a generalized Gini coefficient.
For the ODIAC and FFDAS data it depends on the socio-economic development of the considered country (developing countries exhibit relatively more emissions in locations of high population).
For the EDGAR data there is no development dependence (overall relatively more emissions in locations of low population).
Within China, the development dependence persists for the ODIAC data, but within the USA it vanishes for the ODIAC and Vulcan data.
Sub-sampling and shuffling supports the robustness of our analysis.

The difficulty in explaining the observed phenomenon of country-specific inhomogeneity indices may be attributable to a complex interplay of human activity on local, country, and international scale which entails more evaluation.
Concentration or dispersion of human activities is strongly linked to the extent of urban sprawl.
Such structural properties certainly affect both the population and the emissions.
Moreover, as mentioned earlier, the proxies used to downscale national level CO\textsubscript{2} emissions and the sectors included while calculating the national level emission data will also impact the spatial inhomogeneity of the origin of CO\textsubscript{2} emissions.
In addition, the location of point sources is an important aspect that can hardly be generalized on the national or even international scale.
Maybe, a starting point could be a better understanding of the spatial characteristics of CO\textsubscript{2} efficiency.
Explaining the presented phenomenon -- i.e.\ development dependent concentration of emissions in locations of high or low population -- remains a challenge for future research.

There is a strong association between urbanization, economic development, and carbon emissions.
Here we show that also the location of emissions is influenced by the economic development.
Superposing the structure of the urban texture, the emissions are localized in a development dependent inhomogeneous fashion.
According to our results, with increasing development at a national scale emitting sources shift to less populated areas.
A possible explanation could be an increasing environmental consciousness and adoption of cleaner technologies -- a trend similar to the environmental Kuznets curve (EKC).
While a majority of national mitigation strategies target specific sectors, our results suggest a complementary spatial perspective to prioritize mitigation actions.
Depending on the considered scope of emissions, these would be sparsely populated regions in developed countries and densely populated regions in developing and transition countries.
Particular attentions should be paid to the latter, as these countries are projected to become more urbanized in the upcoming decades, which entails further rural-urban migration.

Our results for the ODIAC and FFDAS data are consistent with previously reported findings \cite{RybskiRWFSK2017}, according to which in developing countries large cities are comparably less efficient in terms of CO\textsubscript{2} emissions, and in developed ones small cities are less efficient. 
On the one hand, the present study provides stronger empirical evidence, e.g.\ because it is based on more data and the signatures are more pronounced.
On the other hand, the methodology of the present study does not rely on any city definition \cite{ArcauteHFYJB2014,RozenfeldRABSM2008} or any assumption about the functional form of the correlations between population and emissions \cite{LeitaoMGA2016}.

We conclude that the affirmation ``large cities are less green'' \cite{OliveiraAM2014} needs to be revised.
According to our results only in developing countries large cities are less green. 
In developed countries, including the USA, the opposite is the case, relatively more emissions stem from small cities.
Anyways, we find it misleading to speak about ``green cities'' in the context of urban CO\textsubscript{2} emissions \cite{GlaeserK2010}, since greenness usually refers to urban vegetation or metaphorically to pollution (while CO\textsubscript{2} is a colorless gas which as a GHG contributes to global warming).

Certainly, our analysis also has some potential caveats which we want to discuss briefly.
The analysis stands and falls with the employed input data, so we cannot exclude to obtain other results if we use other population or emissions data as inputs. 
Why the EDGAR data leads to different results compared to ODIAC and FFDAS is an interesting problem requiring further research.
Moreover, our curves, such as in Fig.~\ref{fig:grump_odiac}, can have (multiple) crossings with the diagonal and the index \egi\ cannot capture to a full extent more complex shapes of the curves.

Another aspect that could be addressed in future studies is the role of the population density \cite{NewmanK1989,GudipudiFRW2016,RybskiRWFSK2017,RybskiD2016}. 
Here we avoid any discussion about city definitions by simply taking gridded data.
Since the grid cells are approximately of equal area, the population count and the density are approximately identical.
In order to investigate the influence of the density, a suitable city definition -- joining grid-cells -- will be necessary.

\section*{Acknowledgment}
We thank M. Barthelemy for useful discussions.
The research leading to these results has received funding from the European Community's Seventh Framework Programme under Grant Agreement 308497 (Project RAMSES). 
Author BZ thanks Climate-KIC, the climate innovation initiative of the EU's European Institute of Innovation and Technology (EIT), for award of a Ph.D.\ scholarship.
ODIAC emissions dataset was provided by T. Oda of Colorado State University, Fort Collins CO, USA/Global Monitoring Division, NOAA Earth System Research Laboratory, Boulder CO, USA. Odiac project is supported by Greenhouse Gas Observing SATellite (GOSAT) project, National Institute for Environmental Studies (NIES), Japan.

\section*{References}
\bibliographystyle{jphysicsB}
%\bibliography{egbib}

\clearpage

\appendix

\section{Additional Figures}
\label{sec:subnatusa}

\begin{figure}
\includegraphics[width=\linewidth]{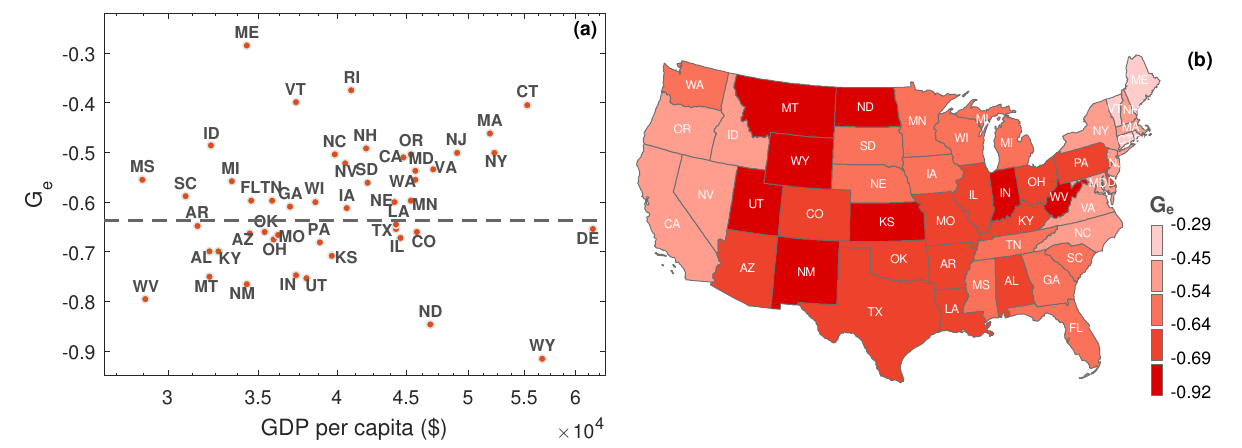}
\caption{Sub-national inhomogeneity index \egi{}. We calculated the \egi{} on the state level for the USA. In (a) the \egi{}-values are plotted against the corresponding state GDP per capita values on a logarithmic scale (excluding District of Columbia), analogous to Fig.~\ref{fig:ge_vs_gdp}.
The dashed line indicates the country-level mean \egi.
Panel (b) shows a map of contiguous USA where the states are color-coded according to the inhomogeneity index \egi{}. 
It can be seen that the development dependence as found in Fig.~\ref{fig:ge_vs_gdp} does not hold on the sub-national scale 
-- at least for the USA.
However, spatially the values are not random, and we find large \egi{}-values at the east and west coasts and smaller values in the predominantly sparsely populated states.}
\label{fig:usa}
\end{figure}

In Fig.~\ref{fig:usa}(a) the \egi{}-values are plotted vs.\ the corresponding GDP per capita values, as in Fig.~\ref{fig:ge_vs_gdp} but here for states in the USA (analogous to Fig.~\ref{fig:china}).
In contrast to the country analysis, we do not find correlations ($\rho=0.07$, p-value: $0.64$, not statistically significant). 
However, the \egi{}-values are consistently in the negative range so that overall high population densities come along with lower CO\textsubscript{2} per capita (consistent with Fig.~\ref{fig:ge_vs_gdp}).

\begin{figure}
\includegraphics[width=\linewidth]{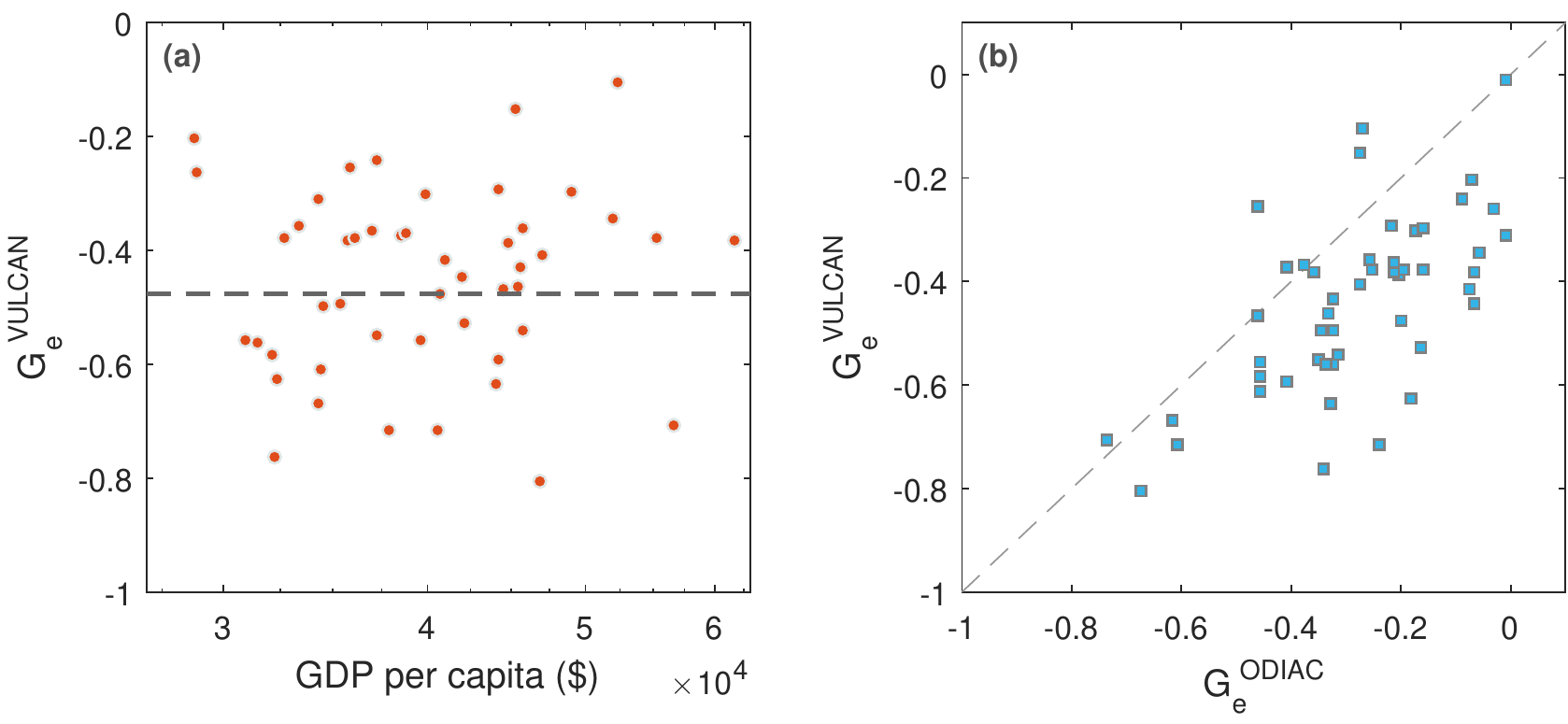}
\caption{
Sub-national inhomogeneity index \egi{}. We calculated the \egi{} on the state level for the USA based on th Vulcan data for the year 2002 at 10\,km resolution \cite{GurneyMZF2009,GurneyRSZ2012} . 
In (a) the \egi{}-values are plotted against the corresponding state GDP per capita values on a logarithmic scale (excluding District of Columbia), analogous to Fig.~\ref{fig:usa}(a).
The dashed line indicates the country-level mean \egi.
It can be seen that also for the Vulcan data the development dependence does not hold on the sub-national scale in the USA.
In (b) we show the correlations between the \egi{} obtained from the ODIAC data and the corresponding values obtained from the Vulcan data.}
\label{fig:usavulcan}
\end{figure}

Results of the analogous analysis for the USA and the Vulcan data are displayed in Fig.~\ref{fig:usavulcan}(a).
As can be seen, still there are no correlations between the obtained \egi{}-values and the GDP per capita.
Comparing the resulting \egi{}-values from the Vulcan data with those obtained for the ODIAC data, we do find weak correlations [Fig.~\ref{fig:usavulcan}(b)].
In comparison to the ODIAC, the Vulcan data overall tends to exhibit lower \egi{}-values, indicating that there are more emissions from sites of low population.

\pagebreak
\section{Derivation of the relationship between $\beta$ and \egi}
\label{sec:betagini}
Denoting probability distribution functions with $F$, the theoretical quasi-Lorenz curve for emissions $E\sim F_E$ with respect to population $P\sim F_P$ is defined as
\begin{equation}
L_{E \circ P}(\theta) = \frac{1}{\mu_E}\int\limits_{-\infty}^{S_P^{-1}(\theta)}\mu_{E|P}(t)\dif F_P(t) \quad 0\le\theta\le 1
\end{equation}
where $\mu_P$ and $\mu_E$ are the respective means of $P$ and $E$ and $\mu_{E|P}$ is the conditional mean of $E$ given $P$. In contrast to the classical concentration curves \cite{Yitzhaki1991}, the upper boundary of integration is given through the generalized inverse of $S_P(p)$
\begin{equation}
S_P^{-1}(\theta)=\inf\{p: S_P(p)\ge \theta\}.
\end{equation}
We call $S_P(p)$ the share function defined as
\begin{equation}
S_P(p)=\frac{1}{\mu_P}\int\limits_{-\infty}^p t \dif F_P(t).
\end{equation}
If we assume that the population $P$ is Pareto distributed with shape parameter $\lambda >1$ and scale $p_\mathrm{min}>0$, the inverse share function $S_P^{-1}(\theta)$ is given through
\begin{equation}
S_P^{-1}(\theta)=p_\mathrm{min}(1-\theta)^{\frac{1}{1-\lambda}}.
\end{equation}
If we further assume that the scaling relation $E=aP^\beta$ holds, the conditional mean is simply given as $\mu_{E|P}(t)=a t^\beta$ and the unconditional mean for $\beta< \lambda$ can be calculated as
\begin{equation}
\mu_Y = \frac{\lambda}{\lambda-\beta} a p_\mathrm{min}^\beta.
\end{equation}
If $\beta\geq\lambda$ the unconditional mean becomes infinite and the quasi-Lorenz curve can not be computed.
Given the previous assumptions the quasi-Lorenz curve can be derived as
\begin{equation}
L_{E \circ P}(\theta)=\left[\frac{\lambda}{\lambda-\beta} a p_\mathrm{min}^\beta\right]^{-1}\int\limits_{p_\mathrm{min}}^{p_\mathrm{min}(1-\theta)^{\frac{1}{1-\lambda}}} a \lambda p_\mathrm{min}^\lambda t^{\beta-\lambda-1} \mathrm{d}t
\end{equation}
which simplyfies to
\begin{equation}
L_{E \circ P}(\theta)= 1- \left(1-\theta\right)^{\frac{\lambda-\beta}{\lambda-1}}.
\end{equation}
The generalized Gini coefficient \egi{} is then given by
\begin{equation}
G_e =  1 - 2 \int\limits_0^1 L_{E \circ P}(\theta) \mathrm{d}\theta = \frac{\beta-1}{2\lambda-\beta-1}
\end{equation}
as stated in Eq.~(\ref{equ:betagini}).

\end{document}